\documentstyle[12pt]{article}

\begin{document}
%format latex
\title{Crossover from Fermi Liquid to Non-Fermi Liquid Behavior in a
Solvable One-Dimensional  Model}
\author{Y. Chen and K. A. Muttalib$^{\dag}$\\
Department of Mathematics\\
Imperial College, London SW7 2BZ, U. K.}
\maketitle
\begin{abstract}
We consider a quantum many-body problem in one-dimension described by a
Jastrow type wavefunction, characterized by an exponent $\lambda$ and
a parameter $\gamma$. In the limit $\gamma=0$ the model
becomes identical to the well known $1/r^2$ pair-potential model;
$\gamma$ is shown to be related to the strength of a many body
correction to the $1/r^2$ interaction. Exact results for the
one-particle density matrix are obtained for all $\gamma$ when
$\lambda=1$, for which the  $1/r^2$ part of the interaction vanishes. We
show that with increasing  $\gamma$, the Fermi liquid state (at
$\gamma=0$) crosses over to distinct $\gamma$-dependent non-Fermi liquid
states, characterized by  effective ``temperatures''.
\vskip 1cm
\par\noindent
$^{\dag}$Permanent address: Physics Department, University of Florida,
Gainesville, FL 32611, U.S.A.
\end{abstract}
\newpage
A special class of Jastrow type wave functions  [1]
\begin{equation}
\Psi(u_1,\ldots,u_N)=C\prod_{1\leq a<b\leq N}|u_a-u_b|^{\lambda}
\;\prod_{c=1}^{N}{\rm e}^{-V(u_c)},\end{equation}
appear frequently in quantum many-body problems. In one dimension,
many-body Hamiltonians with pair-potentials
of the form $1/r^2$ (or its periodic equivalent $1/{\rm sin}^2r$), where
 $r=(u_a-u_b)$,
have exact ground state wave functions of the Jastrow form [2-5]. It has
been proposed that variational
wave functions of this type give reasonably good descriptions of
models for strongly interacting fermions [6-11]. It is therefore of
interest to investigate
the properties of such wave functions as exactly as possible. Of
particular interest is the question whether such a wave function
describes how interaction can change a Fermi liquid into a non Fermi
liquid state.
\par\indent
In the solution of the $1/r^2$ model, the parameter $\lambda$ in Eq. (1)
is related to the strength of the $1/r^2$ pair potential; in particular,
$\lambda=1$ corresponds to the free fermion case. Unfortunately, the
exact density matrix for this model can be obtained  only for a few
special values of $\lambda$  [2,12], so the question of the nature of
crossover from the free fermion to the interacting non-Fermi liquid
state can not be addressed exactly in this model. In the present work,
we consider a one-dimensional wave function of the above form which can
be considered as a
generalization of the wavefunction corresponding to the $1/r^2$ model.
In addition to the parameter $\lambda$, our model contains an additional
parameter $\gamma$ which we show to be related to the strength of a many
body correction to the $1/r^2$ interaction. We obtain the
one-particle density matrix for this model exactly for all $\gamma$, for
the particular case of $\lambda=1$. Thus $\gamma=0$ represents the free
fermion case. We show that with increasing  $\gamma$, the Fermi liquid
state is destroyed by an effective non-zero ``temperature'' induced by
interaction.
\par\indent
The exact density matrix of the $1/r^2$ model was obtained [2] by
exploiting the analogy of the wave function with the eigenvalue
distribution of random matrices [13]. The wavefunction we consider is
motivated by our recent generalization of the conventional
Wigner-Dyson-Mehta random matrices, for describing transport in
disordered systems [14]. We consider the wavefunction given by Eq.(1)
with
\begin{equation}
V(u)={1\over 2\gamma}\left[{\rm sinh}^{-1}[(\gamma
\omega)^{1\over 2} u]\right]^{2}
+{1\over 2}{\rm ln}
\left[\vartheta_4\left({\pi\over \gamma}
{\rm sinh}^{-1}[(\gamma
\omega)^{1\over 2} u];p\right)\right],
\end{equation}
characterized by a single parameter $\gamma$. Here $\vartheta_4(x;p)$ is
the
Jacobi Theta function [15], $p={\rm e}^{-\pi^2/\gamma}$, and $\omega$
has the
dimension of $1/[{\rm length}]^2$.
For $\gamma=0,$ $V(u)=
{1\over 2}\omega u^2,$
and the wavefunction reduces to the well known solution of
the $1/r^2$ pair potential. The case $\lambda=1$ then represents a free
Fermion
problem (with the choice of Fermi statistics), with the one-particle
density matrix given by
${{\rm sin}[\pi Dr]/(\pi r)},$ $D$ being the density of particles. The
parameter
$\gamma$ ``deforms'' the
harmonic well into a weakly confining $[{\rm ln}u]^2$ term for large
enough $u$, as shown in figure 1.
We will show that this deformation leads to a qualitative change in the
density matrix; for $\lambda=1$ this corresponds to a change from
a Fermi liquid to a non-Fermi liquid state.
\par\indent
In order to understand the role of the parameter $\gamma$, let us
concentrate for the moment on the small $\gamma (\ll \pi^2)$ limit where
the second term in Eq. (2) can be neglected, and the Hamiltonian
corresponding to the above wave function has a simple form. In this
limit, to leading order in $\gamma$, the Schr\"odinger
equation (in units where $\hbar^2/2m=1$)
can be written as,
$$
{1\over \Psi}\sum_{k}{\partial^2\over \partial u_k^2}\Psi
=\lambda(\lambda-1)\sum_{j\neq k}{1\over (u_k-u_j)^2}
$$
$$
+{\omega}^2[1+2\gamma]\sum_k{(u_k)^2}
-\omega[N+\lambda N(N-1)]$$
\begin{equation}
+{2\over 3}\lambda\gamma{\omega}^2\sum_{j\neq k}{1\over
u_k-u_j}[(u_k)^3-(u_j)^3]
-{4\over 3}\gamma{\omega}^3\sum_k{(u_k)^4}
\end{equation}
Note that for $\gamma=0$, the Hamiltonian reduces to
the $1/(u_k-u_j)^2$ pair potential, with energy $E=\omega[N+\lambda
N(N-1)]$ [2,3].
For $\lambda=1,$ with the choice of Fermi statistics,
this becomes a free Fermion problem.
However, for $\gamma\neq 0$, a many-body correction term survives even
for
$\lambda=1$. Nevertheless,
the density matrix for $\lambda=1$ can still be
found exactly for all $\gamma$. We will show that it corresponds to a
non-Fermi liquid state, $\gamma$ playing the role of an effective
temperature.
\par\indent
The exact density matrix corresponding to the wave function defined by
eqns. (1) and (2) for  $N\rightarrow \infty$ and for all $\gamma$, as
obtained by exploiting the analogy of the present problem with
the random matrix model recently constructed for describing transport in
disordered systems [14], is given by
\begin{equation}
\rho(u,v)=f(\gamma){\cal Q}(\mu,\nu)
{\vartheta_1\left({\pi(\mu-\nu)\over 2\gamma};p\right)\over u-v},
\end{equation}
where
\begin{equation}
{\cal Q}(\mu,\nu)={\vartheta_4\left({\pi(\mu+\nu)\over 2\gamma};p\right)
\over \vartheta_4^{1/2}\left(\frac{\pi\mu}{\gamma};p\right)
\vartheta_4^{1/2}\left({\pi\nu\over \gamma};p\right)},\;
\mu={\rm sinh}^{-1}(\sqrt {\gamma
\omega} u),\;\;\;\;\nu={\rm sinh}^{-1}(\sqrt {\gamma
\omega} v),
\end{equation}
$\vartheta_1(x;p)$ is a Jacobi Theta function [15]
and
$f(\gamma)$ is a known function of $\gamma$.
For $\gamma\ll \pi^2,$ a simpler form for the density matrix is obtained
\begin{equation}
\rho(u,v)\approx {1\over \pi}
{{\rm sin}\left[{\pi\over 2\gamma}\left(
{\rm sinh}^{-1}\left[(\omega\gamma)^{1/2}u\right]
-{\rm sinh}^{-1}\left[(\omega\gamma)^{1/2}v\right]\right)\right]\over
u-v}.
\end{equation}
In the limit $\gamma\rightarrow 0$, in terms of the density at the
origin $D_0=\frac{1}{2}(\frac{\omega}{\gamma})^{1/2},$
this reduces to the free Fermion density matrix
$\rho(u-v)={{\rm sin}[D_0\pi(u-v)]/\pi(u-v)}.$
These oscillations are the characteristic signature of a normal
Fermi system; its Fourier transform---the momentum distribution---
is the familiar step function. On the other hand for increasing
$\gamma$,
these oscillations begin to die out, destroying the Fermi liquid
behavior. The Fourier
transform of $\rho(u,v)$ would involve two external momenta and this
makes
comparison with the Fermi liquid  momentum distribution difficult.
We shall instead consider the Fourier transform of
$\rho(u,0),$ $n_k=\int_{-\infty}^{+\infty}du\; {\rm e}^{iku}\rho(u,0)$.
With
the change of variable $y={\rm sinh}^{-1}(2\gamma D_0 u),$ we have
\begin{equation}
n_k=\frac{1}{\pi}\int_{0}^{\infty}dy\;{\rm coth}y
\left[{\rm sin}\frac{\pi}{2\gamma}\left(\frac{|k|}{\pi D_0}{\rm
sinh}y+y\right)
-{\rm sin}\frac{\pi}{2\gamma}\left(\frac{|k|}{\pi D_0}
{\rm sinh}y-y\right)\right].
\end{equation}
For small enough $\gamma$ we can replace ${\rm sinh}y$ by $y$ within the
$sine$ function, and ${\rm coth}y$ by $1/{\rm sinh}y$. The resulting
distribution is given by
\begin{equation}
n_{k}=n\left[\frac{\pi}{\gamma}\left(\frac{|k|}{D_0}-\pi\right)\right]-
n\left[\frac{\pi}{\gamma}\left(\frac{|k|}{D_0}+\pi\right)\right],
\end{equation}
where $n[x]=1/[{\rm e}^x+1]$ is the Fermi function.
As expected, this reduces to the step of the free fermions when
$\gamma=0.$ Moreover,
we find from the explicit expression (8) that $\gamma$ plays the role of
an effective ``temperature''.
\par\indent
In summary, we have considered a one-dimensional quantum many-body
problem
described by a Jastrow type wave function characterized by an exponent
$\lambda$
and a parameter $\gamma$. We show that our model is a generalization of
the $1/r^2$ pair-potential model considered by Calogero and Sutherland,
which is obtained in the limit $\gamma=0.$ We obtain the exact
one-particle density matrix for all $\gamma$ for the case $\lambda=1$
where the $1/r^2$ interaction vanishes. For $\gamma=0$, and the choice
of Fermi statistics
this becomes a free fermion problem, and we recover the step function
for the
momentum distribution. For $\gamma\neq 0$, an interaction term survives
for
the case $\lambda=1$ and the resulting momentum distribution is smeared
out,
destroying the Fermi Liquid. The explicit expression for the momentum
distribution
as a function of $\gamma$ for small $\gamma$ shows that the destruction
of the Fermi liquid state occurs as increasing interaction induces an
increase in the effective ``temperature'' in this regime.
\par\indent
One of us (K. A. M.) should like to thank the Science and Engineering
Research
Council, UK, for the award of a Visiting Fellowship, and the Mathematics
Department
of Imperial College for its kind hospitality. We should also like to
thank
A. C. Hewson for discussion, and J. R. Klauder for valuable comments on
the manuscript.

\vskip 4cm
{\bf Figure Caption}
\par\noindent
Fig.1: $V(u)$ characterizing the model wave function considered, as given
by Eq.(2), for various values of $\gamma$. The parameter $\omega$ has
been set equal to unity.

\par\noindent
\end{document}